\documentclass{article}
\usepackage{spconf,amsmath,graphicx}

\usepackage{breqn}
\usepackage{subfigure}
\usepackage{xcolor}

\usepackage{amssymb}

\begin{document}

\title{A One-Shot Learning Framework for Assessment of Fibrillar Collagen from Second Harmonic Generation Images of an Infarcted Myocardium}

\name{ \begin{tabular}{c}
Qun Liu \textsuperscript{1},
Supratik Mukhopadhyay \textsuperscript{1},
Maria Ximena Bastidas Rodriguez \sthanks{Universidad Nacional de Colombia, Bogota, Colombia.},
Xing Fu \textsuperscript{1}, \\
Sushant Sahu \textsuperscript{1},
David Burk \textsuperscript{1},
Manas Gartia \textsuperscript{1}\end{tabular} \vspace{-8pt}}

\address{\textsuperscript{1} Louisiana State University, Baton Rouge, LA 70803, USA \\
\{qliu14, ssahu, mgartia\}@lsu.edu, supratik@csc.lsu.edu, \\ xfu1@agcenter.lsu.edu, david.burk@pbrc.edu \vspace{-6pt}}

\maketitle

\begin{abstract}
Myocardial infarction (MI) is a scientific term that refers to heart attack.
In this study, we infer highly relevant  second harmonic generation (SHG) cues from collagen fibers exhibiting  highly  non-centrosymmetric assembly  together with two-photon excited cellular autofluorescence in infarcted mouse heart to quantitatively probe fibrosis, especially targeted at an early stage after MI. We present  a robust   one-shot machine learning  algorithm that enables determination of  2D assembly  of collagen  with high spatial resolution along with its  structural arrangement in heart tissues post-MI with spectral specificity and sensitivity. Detection, evaluation, and precise quantification of fibrosis extent at early stage would guide one to develop treatment therapies that may prevent further progression and  determine heart transplant needs for patient survival. 

\end{abstract}
\vspace{-6pt}
\section{Introduction}
\vspace{-5pt}
Myocardial infarction (MI) arises due to lasting damage suffered by cardiomyocytes as a result of  blockage in blood circulation over an extended period of time   to an area within the heart. In the course of the next few weeks following MI, permanently damaged  cardiomyocytes are slowly restored  by collagen containing  scar tissue mainly comprising of fibroblasts and extracellular matrix (ECM) proteins because of the lack of cardiomyocyte regeneration~\cite{3}. 
The passage of   myocardial scar  repair  post infarction is time-dependent and is  comprised of three different phases, namely,  inflammation/necrosis, fibrosis/proliferation, and continued left ventricular  remodeling/maturation \cite{3}. 
Improper function  of left ventricle due to MI and  left ventricular remodeling  can result in  cardiac failure post MI~\cite{3}.

The increased collagen contents \cite{9} in fibrosis is traditionally analyzed with tedious histological and immunohistochemical methods which needs irreversible  isolation of cardiac tissue specimens, dissection into thin sections using microtomes (1-10 $\mu$m), embedding, fixation, and staining procedures. Moreover, acquiring myocardial tissue images in the clinical environments is difficult, as pretreatment procedures acting on sample processing can not only  corrupt but also can  induce structural modifications in tissue specimens. 

Biopsy has been  the gold standard for myocardial fibrosis. Manual chemical staining and histopathology of these biopsy slides often lead to inter-observer variability, inadequate quantification of accuracy, and lack of high-confidence in the data. Another problem with large volume of samples is the labor cost and time of analysis. In order to reduce time and cost, pathologists often utilize partial sampling strategy where instead of analyzing all the slides, only few randomly chosen slides are analyzed. Since, the diagnosis and therapy rely on these histopathological assessments, the current procedures lead to inconsistent, inaccurate, and unreliable results leading to unfavorable patient outcomes \cite{ieee}. Hence, there is an unmet need to develop an automated technique to perform label free assessment of cardiac tissues using their inherent molecular contrast. 

\begin{figure*}[t!]
\centering
\includegraphics[width=0.75\textwidth]{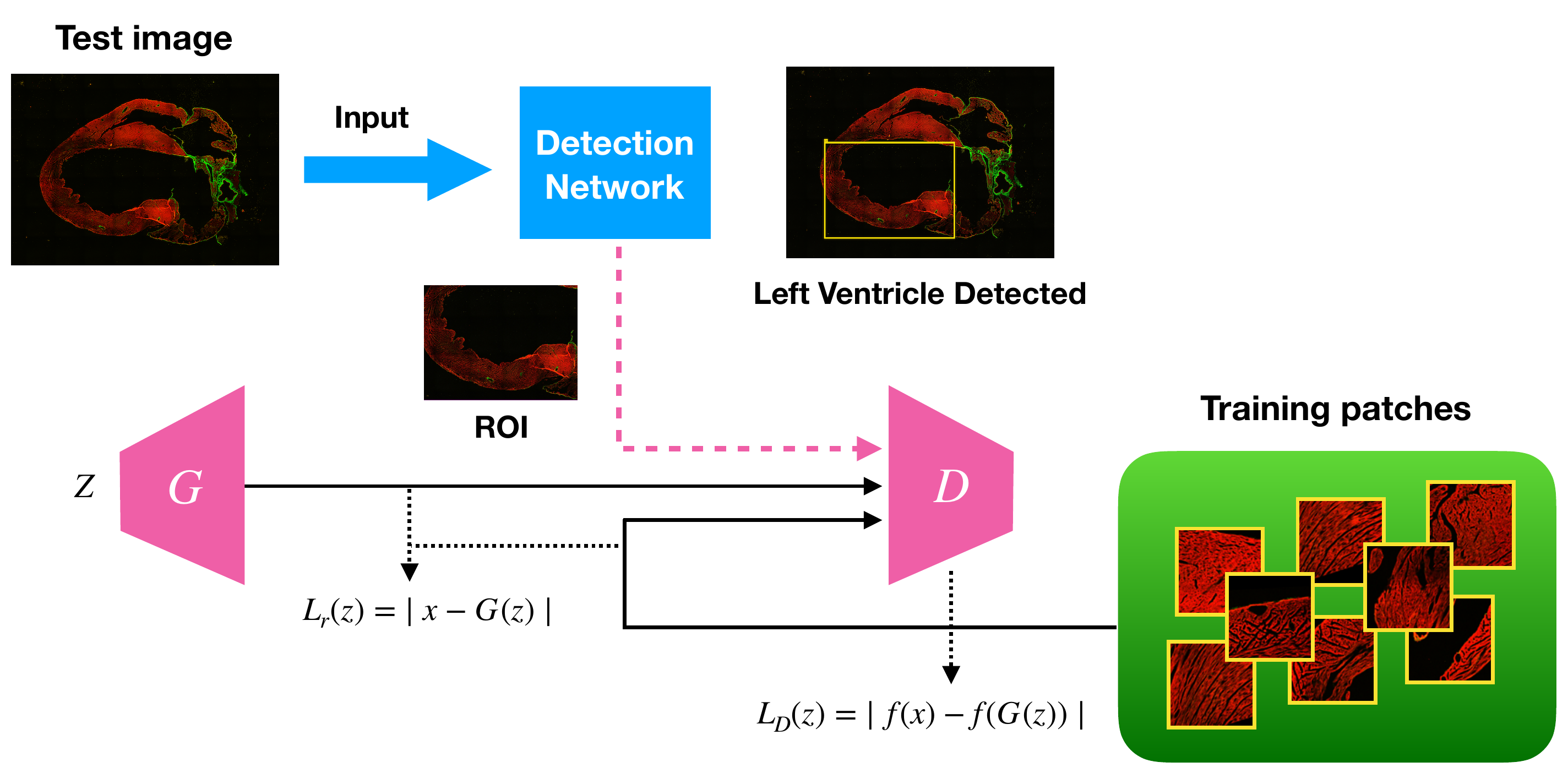}
\vspace{-5pt}
\caption{Architecture of our proposed framework.}
\label{fig:bf}
\vspace{-6pt}
\end{figure*}

 Recent years have seen the rise of non-linear optical microscopy techniques like second harmonic generation (SHG) microscopy \cite{17} as  powerful techniques for non-invasive imaging of extracellular matrix (ECM) and cellular structures \cite{17} of thick tissue specimens in their native environments~\cite{17}. Second Harmonic Generation (SHG) \cite{17} prvides a coherent second order nonlinear optical scattering method usually by molecules  having asymmetric structures  which allows conversion of two lower energy photons into single photon with twice the incident energy of an excitation laser light. The SHG photon is nonisotropic  hence polarization-sensitive and can be  synthesized  within  femto-seconds. This allows using  femtosecond pulsed lasers, to produce  SHG signals with comparable  reaction  time scales. In contrast to conventional confocal imaging penetration depth ($\sim 60 \: \mu$ m), SHG has higher penetration depth ($\sim 600 \: \mu$ m) to image thick tissue specimens and intrinsic optical sectioning capability~\cite{18}.
SHG microscopy is  a label-free technique for imaging native structure of tissue specimens without any requirement for staining, thus preventing any corruption due to staining~\cite{17}.

In this study, we infer highly relevant  second harmonic generation (SHG) cues from collagen fibers exhibiting  highly  non-centrosymmetric assembly  together with two-photon excited cellular autofluorescence in infarcted mouse heart to quantitatively probe fibrosis, especially targeted at an early stage after MI. Employing a computational framework comprising robust   one-shot machine learning  algorithms,   we determine  2D assembly  of collagen  with high spatial resolution along with its  structural arrangement in heart tissues post-MI with spectral specificity and sensitivity (we limit this study to 2D as obtaining high resolution 3D SHG images through collection of Z-stacks of images on whole mount or thick tissue specimens would require lot of time and resources). More precisely, the  framework uses a detection network for localizing  the left ventricle in an SHG microscopy image of a heart and   a  generative  adversarial network (GAN) \cite{GAN1}   to  infer a segmentation mask for  the infarction area (in the left ventricle), by estimating the heat map representing the anomaly   between a single SHG microscopy image of a   normal heart and that of an infarcted one. The segmented infarction area is used to compute an infarction score to quantitatively assess the fibrosis extent. Detection, evaluation, and precise quantification of fibrosis extent at early stage would guide one to develop treatment therapies that may prevent further progression and heart transplant needs for patient survival.

\begin{figure*}[t!]
\centering
\includegraphics[width=0.5\textwidth]{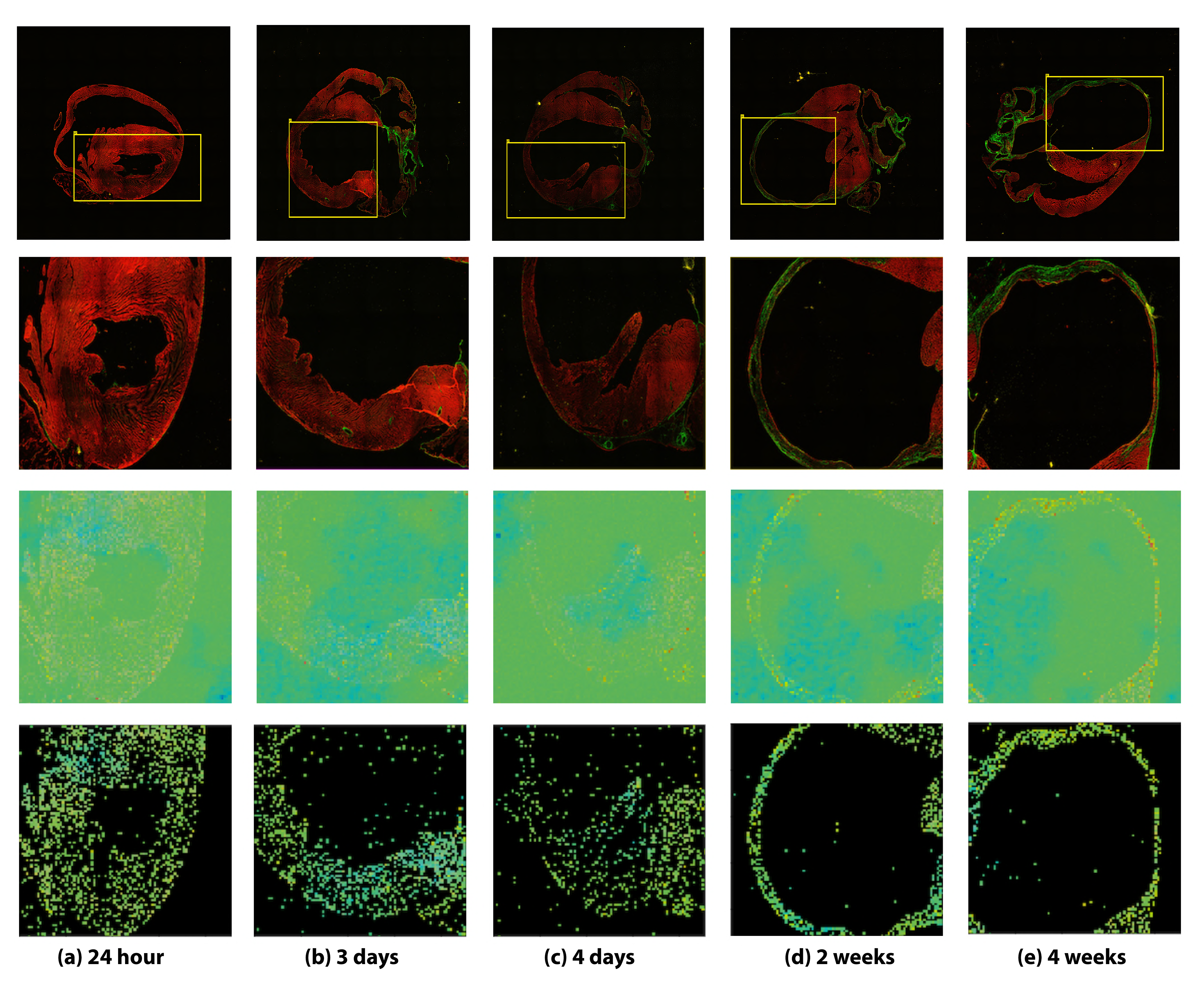}
\vspace{-10pt}
\caption{Segmentation of infarcted region in left ventricle. The top row shows the detection of the region of interest (left ventricle) using detection network. The second row presents the left ventricular region. Third row exhibits the residual difference heat map for infarction area computed as segmentation mask from  the GAN. And the bottom row, shows the segmentation of the infarction area based on the segmentation mask.}
\label{fig:5result}
\vspace{-6pt}
\end{figure*}

\vspace{-10pt}
\section{Methods}
\vspace{-5pt}
\subsection{Framework Overview}
\vspace{-5pt}
Fig. \ref{fig:bf} presents the architecture for the one-shot machine learning and image processing pipeline that constitutes the basis of our framework. The pipeline  consists of two components: a detection network for localizing  the left ventricle in an SHG microscopy image of a heart  and a GAN   to  learn a 
segmentation mask for  the infarction area (in the left ventricle), by estimating the heat map representing the anomaly~\cite{26}  between an SHG microscopy image of a normal heart and that of an infarcted one.  A GAN ~\cite{GAN1, liu2019pcgan}  trains two models simultaneously. The first model $G$, called the \emph{generator},  generates samples by attempting to capture the data distribution of the training set (patches from \emph{a single}  SHG microscopy image of a normal, i.e., non-infarcted,  heart); and the second model $D$, called the \emph{discriminator}, discriminates between a  created sample by $G$ and ``normal" images from the training distribution.  Finally, we use the segmented infarction area to compute the infarction score, a measure that we have proposed (see below) to quantitatively assess the heart risk due to the infarction. 

Segmenting the infarction area in the SHG microscopy image of an  infarcted heart is challenging due to the lack of training data.   
Standard deep learning-based  segmentation methods  \cite{badrinarayanan2017segnet,core,collier}
require large volumes of labeled training data and hence are unsuitable for the present purpose. 
Hence, we adopted an unsupervised learning pipeline that uses a GAN  to  efficiently learn  features  for computing the segmentation mask for  the infarction area, by estimating the heat map representing the anomaly   between an SHG microscopy image of a normal heart and that of an infarcted one. Existing threshold-based segmentation techniques cannot automatically estimate the anomaly between the normal heart and that of an infarcted one to quantify the risk of a cardiac attack.   


\vspace{-10pt}
\subsection{Detection Network}
\vspace{-5pt}
It is known that infarction in the left ventricle is   critical with respect to quantitative assessment of heart risk \cite{7}. From a raw SHG microscopy image of a heart,  we first use a detection network to determine the region of interest (ROI) focusing on the left ventricle. For the detection network, we adopted a pretrained (on the Coco dataset) object detection framework, known as YOLO~\cite{YOLO}. 
The detection network  computes the ROI that will be used for segmentation of the infarcted area.

\vspace{-6pt}
\subsection{GAN-based Segmentation Mask Estimation}
\vspace{-5pt}
Patches from the SHG microscopy image of \emph{a  single normal}, i.e., non-infarcted,  heart are used as training data for GAN-based segmentation mask estimation. The SHG microscopy images of infarcted heart collected at different timepoints  were used  for testing. The  GAN  architecture of our framework is similar to that in~\cite{26}.  Let $\mathit{Mic}$  be the set of $N$ SHG microscopy images of a heart collected at different timepoints (before and after surgical ligation; see below), where each image $I_n \in \mathbb{R}^{L \times H}$, $n=1,2,\ldots,N$,  is of size of $l \times h$,   with $L=\{0, \ldots,l-1\}$, $H=\{0, \ldots,h-1\}$, $\mathbb{R}$ being the set of pixel intensities. Let $I \in \mathit{Mic}$ be the SHG microscopy image of a normal (non-infarcted) heart. We randomly extract $M$ image patches of size $s \times s$ from $I$.  These patches constitute our training dataset $X$. 
All input images are resized to a fixed input size of $s \times s$. Being an unsupervised setting, we did not use any labels for training.  The loss function for the GAN consists of two components. 

\textbf{Residual Loss}
For a given patch $x \in X$ input to the discriminator, where $x$ follows the probability distribution $\rho_x$, the generator $G$ will generate an image $G(z)$ with a random noise vector  $z$ (following the distribution $\rho_z$) as input.  We use the residual loss function to ensure that the generated image $G(z)$ is similar to the patch $x$ input to the discriminator. 
The residual loss function~\cite{26} defined by,
$\mathcal{L}_{r} = E_{z \sim \rho_z, x \sim \rho_x} [\mid x - G(z)\mid]$,
where $E$ represents expectation. 

\textbf{Feature Matching Loss}
Similar to \cite{26},  we use the feature matching loss in the feature space of discriminator without using labels in an unsupervised setting. The discriminator works as a feature extractor rather than a classifier. The  feature matching loss defined by,
$\mathcal{L}_f = E_{x \sim \rho_x,z \sim \rho_z} [\mid f(x) - f(G(z))\mid]$,
where $E$ represents expectation, $f(\cdot)$ represents the feature extractor for the intermediate layer of  the discriminator. Finally, the overall loss for GAN based on both losses, defined by,
$\mathcal{L} = (1-\lambda) \cdot \mathcal{L}_r + \lambda \cdot \mathcal{L}_f$,
where $\lambda \in [0,1]$ (we use $\lambda=0.1$ in the experiments below).

\textbf{Segmentation} Since, in the SHG microscopic image of an infarcted heart, the accumulation of collagen is signalled by the green channel, the segmentation of  the infarction area (in the left ventricle) of the SHG microscopic image of an infarcted heart entails identification of the ``green" pixels within the ROI (designating the left ventricle).  For a predicate $p$ defined over  the set of  pixels $x,y$ in an image $I$, define the function $I^p(x,y)$ that evaluates to $1$ if the predicate $p$ is true on the pixel $(x,y)$, and $0$ otherwise. Let $G_t$ be the generator of the GAN after training converges. For an  image $I$ with ROI $\zeta$, let $I^{\zeta,g}$ (resp $I^{\zeta,r}$) be the mean green (resp. red) channel intensity within $\zeta$. Then for a query image $I_q \in \mathit{Mic}$ with ROI $\zeta_q$ (left ventricle), the  green segmentation mask is given by
\vspace{-14pt}
\[I_{q,g}^{\mathit{seg}}=\{(x,y) \in \zeta_q  \mid I_q^{g_q(x,y)>G_t^{\zeta_{\mathit{gen}},g}(z)}(x,y)=1\}\]
where $g_q(x,y)$ denotes  the green channel value at pixel $(x,y)$ of $I_q$, $\zeta_{\mathit{gen}}$  is the ROI for the  image $G_t(z)$ generated by the generator $G_t$, and $z$ is a noise vector. Similarly, the red segmentation mask for $I_q$ is given by 
\vspace{-5pt}
\[I_{q,r}^{\mathit{seg}}=\{(x,y) \in \zeta_q  \mid I_q^{r_q(x,y)>G_t^{\zeta_{\mathit{gen}},r}(z)}(x,y)=1\}\]
where $r_q(x,y)$ denotes the red channel value at pixel $x,y$ of $I_q$.
\vspace{-5pt}
\subsection{Infarction Score}
\vspace{-5pt}
For assessment of  the risk  associated with an infarcted heart, we propose an Infarction Score. Since, in the SHG microscopic image of an infarcted heart, the accumulation of collagen is signalled by the green channel, the infarction score will measure the proportion  of ``green" pixels in the segmented infarction area. For a query image $I_q$, let $T_g^q =\mid I_{q,g}^{\mathit{seg}}\mid$ and  $T_r^q =\mid I_{q,r}^{\mathit{seg}}\mid$.  Then the infarction score for $I_q$ is given by $S = T_g^q / T_r^q$.


\begin{figure}[t]
\vspace{-7pt}
\centering
\begin{minipage}{.41\textwidth}
  \centering
  \includegraphics[scale=0.34]{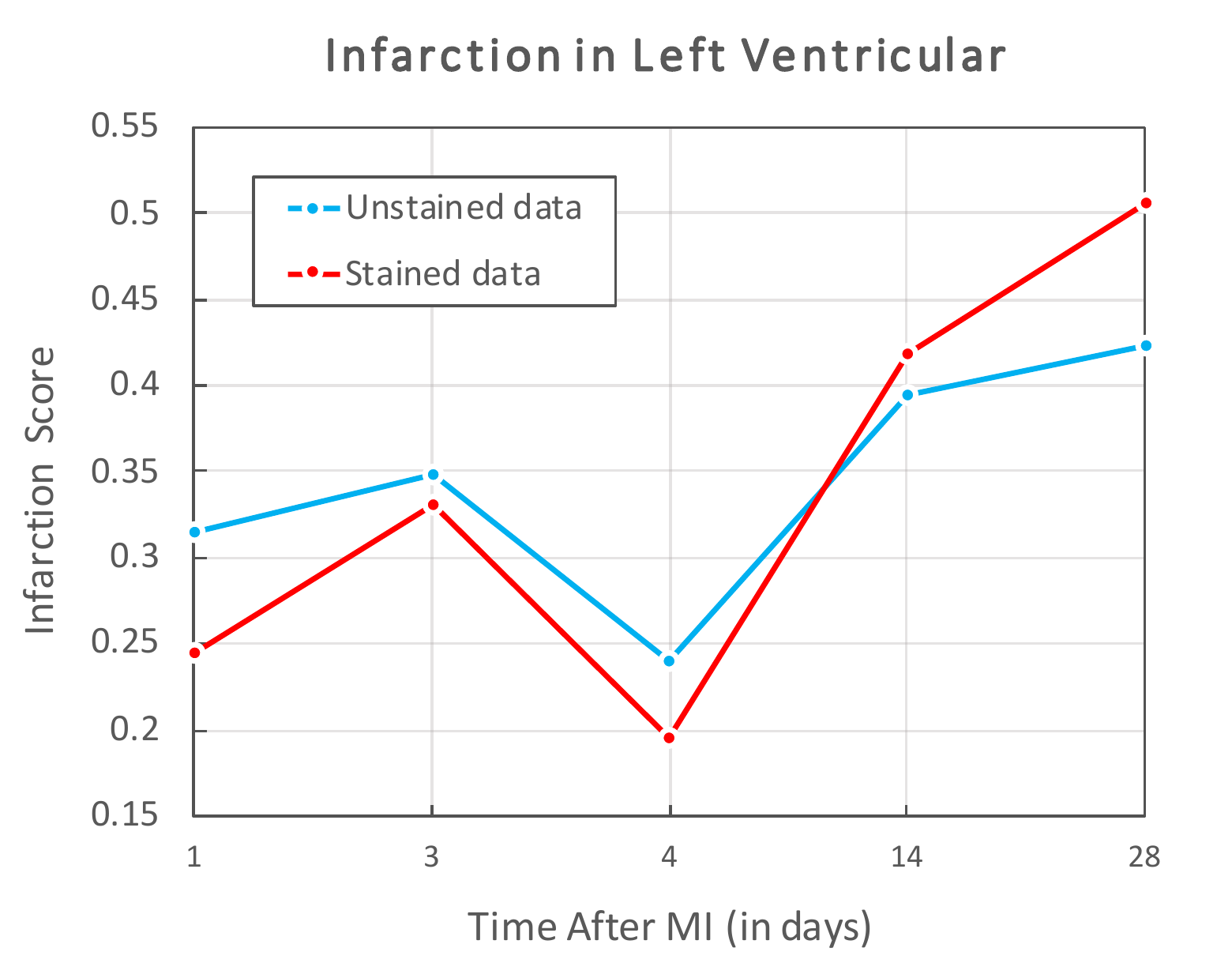}
  \vspace{-6pt}
  \caption{Infarction score corresponding to temporal stages after introduction of myocardial infarction in left ventricle.}
  \label{fig:score}
\end{minipage}%
\hspace{1cm}
\begin{minipage}{.41\textwidth}
  \vspace{15pt}
  \centering
  \includegraphics[scale=0.42]{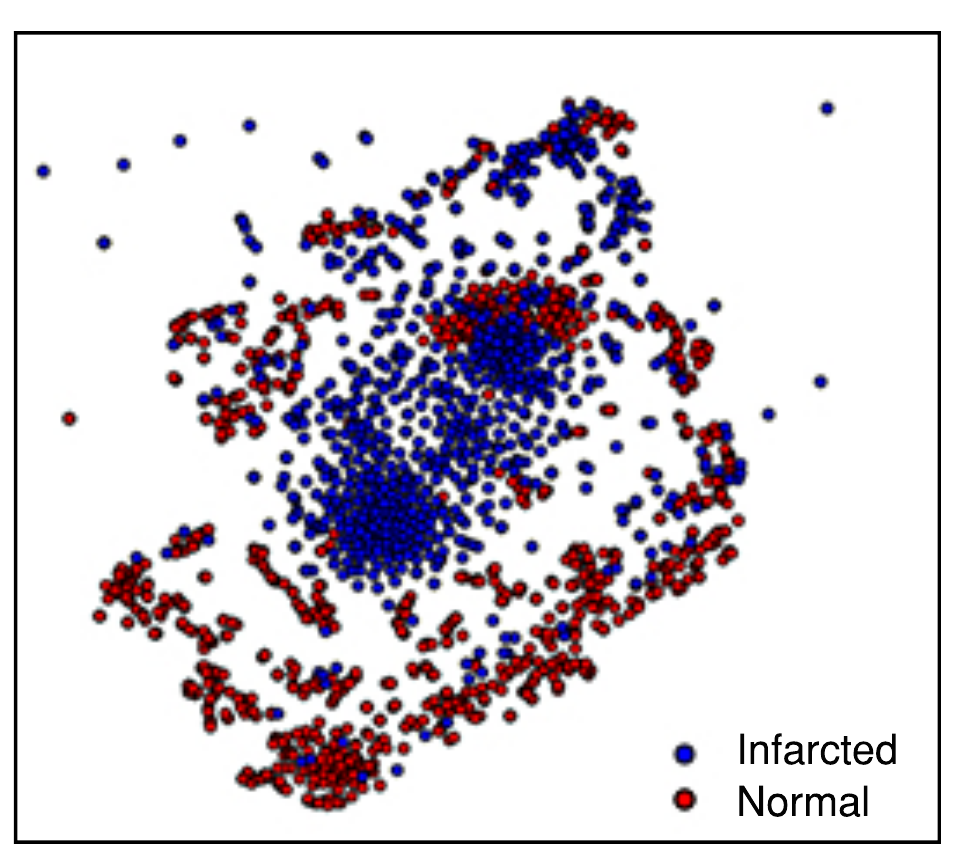}
  \vspace{-6pt}
  \caption{t-SNE visualization of the bottleneck representations from discriminator for infarcted  and normal images.}
  \label{fig:tsne}
\end{minipage}
\end{figure}

\vspace{-10pt}
\section{Experimental Evaluation}
\vspace{-5pt}
In this section, we evaluate our proposed framework on our collected SHG  Heart Imagery Dataset.


\textbf{SHG Heart  Imagery Dataset}
We induced MI  in mice by permanently blocking flow of blood in the left coronary artery using surgical ligation permanent surgical ligation~\cite{21}\footnote{Study received IRB approval from Louisiana State University}. Heart tissue sections were processed and obtained  following protocols published elsewhere~\cite{4}. SHG and Two-photon excited fluorescence imaging was performed using a Leica SP5 resonant scanning multiphoton confocal microscope (Leica Microsystem), coupled to a Spectra Physics Mai-Tai tunable pulsed near-IR laser (690 - 1040 nm) using 70 femtosecond pulses with 84 MHz repetition rate. We carried out all SHG experiments  with an illumination  wavelength of 860 nm. We  collected SHG emission in the epi direction using the 320-430 nm range band pass filters while in case of autofluorescence   486-506 nm bandpass filter was used. After the surgical ligation procedure, SHG microscopic imaging was performed after 24 hours, three days, four days,  two weeks, and four weeks \cite{9,5}. The size of the images were around $7500 \times 6500$ on the average. For separate validation of our MI model through collagen quantification, we analyzed SHG microscopy picture of picrosirius red-stained infarcted mouse heart tissues. The stained images have the same size as that of the unstained ones and were collected for the same timepoints after the surgical ligation procedure. 

\textbf{Experimental Setup}
The generator of the GAN consisted of one dense layer (128 neurons) followed by two conv-transpose layers (64 neurons each).  The discriminator consisted of two convolutional layers (first: 64 neurons, second 128 neurons) followed by a dense layer (1 neuron). We used the bottleneck features of the discriminator. Due to the limited size of our dataset, we only used \emph{a single unstained image} of normal non-infarcted  heart  for training the GAN (trained 500 epochs).   We randomly extracted 1000 patches with patch size of $96 \times 96$ from this training image. When testing, the input test image was be resized to $96 \times 96$ to fit into the input size of the discriminator for computing the mean green channel value for further segmentation.  For picrosirius red-stained images too, we trained the GAN using a \emph{a single stained image} of normal non-infarcted heart.

\textbf{Experimental Results} In  Fig.\ref{fig:5result}, we can see that our proposed method based on one-shot learning shows good performance for the   segmentation task. Fig.\ref{fig:tsne} shows the representations learned by the discrimination. Fig.\ref{fig:tsne} shows  separation between features of SHG images of normal heart (red dots) and those corresponding to images of infarcted heart. Since the  discriminator has learned enriched feature information for normal images, the feature variety results in the red dots scattered throughout the space. Fig.\ref{fig:score} shows the temporal variation in the infarction score after the surgical ligation procedure. Over the course of a week, there have been drastic changes in collagen concentration in the left ventricle. For unstained data (indicated by the blue line in Fig.\ref{fig:score}), on the first day after MI, the infarction score was around 0.31 increasing to 0.34 after three days. However, by 4 days, it had gone down to  0.24. 
By the 28th day however, it increases to 0.41. This is shown in the images as well (see Fig.\ref{fig:5result}).   By the second week, there is a significant portion of green channel. The levels of red (myofibers) start to go down to thinner and thinner levels as well. The results for the stained data (the red line in Fig.\ref{fig:score}) shows  similar trend as that for the  unstained data.


%
\vspace{-8pt}
\section{Conclusion}
\vspace{-5pt}
In this study, we presented a one-shot machine learning approach that 
is able 
 to segment  fibrillar collagens in two-photon excited cellular auto-fluorescence in infarcted mouse heart images in an  unsupervised fashion \cite{unsupervised} using a  deep generative adversarial network. Our framework was able to segment the anomalies presented in the left ventricle due to myocardial infarction  by identifying fibrillar collagens and was able to  assess the risk using an infarction score.



\bibliographystyle{IEEEbib}
\bibliography{main.bib}

\begin{thebibliography}{10}

\bibitem{3}
William~J Richardson, Samantha~A Clarke, T~Alexander Quinn, and Jeffrey~W
  Holmes,
\newblock ``Physiological implications of myocardial scar structure,''
\newblock {\em Comprehensive Physiology}, vol. 5, no. 4, pp. 1877, 2015.

\bibitem{9}
Nagahiro Nishikawa, Tohru Masuyama, Kazuhiro Yamamoto, Yasushi Sakata, Toshiaki
  Mano, Takeshi Miwa, Motoaki Sugawara, and Masatsugu Hori,
\newblock ``Long-term administration of amlodipine prevents decompensation to
  diastolic heart failure in hypertensive rats,''
\newblock {\em Journal of the American College of Cardiology}, vol. 38, no. 5,
  pp. 1539--1545, 2001.

\bibitem{ieee}
S.~Sitharama Iyengar, Xin Li, Huanhuan Xu, Supratik Mukhopadhyay,
  N.~Balakrishnan, Amit Sawant, and Puneeth Iyengar,
\newblock ``Toward more precise radiotherapy treatment of lung tumors,''
\newblock {\em {IEEE} Computer}, vol. 45, no. 1, pp. 59--65, 2012.

\bibitem{17}
Valentina Caorsi, Christopher Toepfer, Markus~B Sikkel, Alexander~R Lyon, Ken
  MacLeod, and Mike~A Ferenczi,
\newblock ``Non-linear optical microscopy sheds light on cardiovascular
  disease,''
\newblock {\em PLoS One}, vol. 8, no. 2, pp. e56136, 2013.

\bibitem{18}
Thomas Abraham, Jeremy~A Hirota, Samuel Wadsworth, and Darryl~A Knight,
\newblock ``Minimally invasive multiphoton and harmonic generation imaging of
  extracellular matrix structures in lung airway and related diseases,''
\newblock {\em Pulmonary pharmacology \& therapeutics}, vol. 24, no. 5, pp.
  487--496, 2011.

\bibitem{GAN1}
Ian Goodfellow, Jean Pouget-Abadie, Mehdi Mirza, Bing Xu, David Warde-Farley,
  Sherjil Ozair, Aaron Courville, and Yoshua Bengio,
\newblock ``Generative adversarial nets,''
\newblock in {\em Advances in Neural Information Processing Systems 27},
  Z.~Ghahramani, M.~Welling, C.~Cortes, N.~D. Lawrence, and K.~Q. Weinberger,
  Eds., pp. 2672--2680. Curran Associates, Inc., 2014.

\bibitem{26}
Thomas Schlegl, Philipp Seeb{\"o}ck, Sebastian~M Waldstein, Ursula
  Schmidt-Erfurth, and Georg Langs,
\newblock ``Unsupervised anomaly detection with generative adversarial networks
  to guide marker discovery,''
\newblock in {\em International Conference on Information Processing in Medical
  Imaging}. Springer, 2017, pp. 146--157.

\bibitem{liu2019pcgan}
Qun Liu, Edward Collier, and Supratik Mukhopadhyay,
\newblock ``Pcgan-char: Progressively trained classifier generative adversarial
  networks for classification of noisy handwritten bangla characters,''
\newblock in {\em International Conference on Asian Digital Libraries}.
  Springer, 2019, pp. 3--15.

\bibitem{badrinarayanan2017segnet}
Vijay Badrinarayanan, Alex Kendall, and Roberto Cipolla,
\newblock ``Segnet: A deep convolutional encoder-decoder architecture for image
  segmentation,''
\newblock {\em IEEE transactions on pattern analysis and machine intelligence},
  vol. 39, no. 12, pp. 2481--2495, 2017.

\bibitem{core}
Manohar Karki, Robert DiBiano, Saikat Basu, and Supratik Mukhopadhyay,
\newblock ``Core sampling framework for pixel classification,''
\newblock in {\em Artificial Neural Networks and Machine Learning - {ICANN}
  2017 - 26th International Conference on Artificial Neural Networks, Alghero,
  Italy, September 11-14, 2017, Proceedings, Part {II}}, 2017, pp. 617--625.

\bibitem{collier}
Edward Collier, Kate Duffy, Sangram Ganguly, Geri Madanguit, Subodh Kalia,
  Shreekant Gayaka, Ramakrishna~R. Nemani, Andrew~R. Michaelis, Shuang Li,
  Auroop~R. Ganguly, and Supratik Mukhopadhyay,
\newblock ``Progressively growing generative adversarial networks for high
  resolution semantic segmentation of satellite images,''
\newblock in {\em 2018 {IEEE} International Conference on Data Mining
  Workshops, {ICDM} Workshops, Singapore, Singapore, November 17-20, 2018},
  2018, pp. 763--769.

\bibitem{7}
Anja~M van~der Laan, Matthias Nahrendorf, and Jan~J Piek,
\newblock ``Healing and adverse remodelling after acute myocardial infarction:
  role of the cellular immune response,''
\newblock {\em Heart}, vol. 98, no. 18, pp. 1384--1390, 2012.

\bibitem{YOLO}
Joseph Redmon, Santosh Divvala, Ross Girshick, and Ali Farhadi,
\newblock ``Youonly look once: Unified, real-time object detection,''
\newblock {\em arXivpreprint arXiv:1506.02640}, 2015.

\bibitem{21}
Jop~H Van~Berlo, Onur Kanisicak, Marjorie Maillet, Ronald~J Vagnozzi, Jason
  Karch, Suh-Chin~J Lin, Ryan~C Middleton, Eduardo Marb{\'a}n, and Jeffery~D
  Molkentin,
\newblock ``C-kit+ cells minimally contribute cardiomyocytes to the heart,''
\newblock {\em Nature}, vol. 509, no. 7500, pp. 337, 2014.

\bibitem{4}
Xing Fu, Hadi Khalil, Onur Kanisicak, Justin~G Boyer, Ronald~J Vagnozzi,
  Bryan~D Maliken, Michelle~A Sargent, Vikram Prasad, I{\~n}igo
  Valiente-Alandi, Burns~C Blaxall, et~al.,
\newblock ``Specialized fibroblast differentiated states underlie scar
  formation in the infarcted mouse heart,''
\newblock {\em The Journal of clinical investigation}, vol. 128, no. 5, 2018.

\bibitem{5}
JP~Cleutjens, MJ~Verluyten, JF~Smiths, and MJ~Daemen,
\newblock ``Collagen remodeling after myocardial infarction in the rat
  heart.,''
\newblock {\em The American journal of pathology}, vol. 147, no. 2, pp. 325,
  1995.

\bibitem{unsupervised}
Qun Liu and Supratik Mukhopadhyay,
\newblock ``Unsupervised learning using pretrained cnn and associative memory
  bank,''
\newblock in {\em 2018 International Joint Conference on Neural Networks
  (IJCNN)}. IEEE, 2018, pp. 01--08.

\end{thebibliography}

\end{document}